\documentclass[10pt,letter,twocolumn]{article}
\usepackage{graphicx}
\usepackage{epsfig}

\usepackage{amssymb}

\usepackage[colorlinks=true, linkcolor=blue, citecolor=blue]{hyperref}

\usepackage{amsfonts} 
\usepackage{mathrsfs}
\usepackage{amssymb}
\usepackage{stmaryrd}
\usepackage{amsmath}
\usepackage{bbm}
\usepackage{array}
\usepackage{float}
\usepackage{subfloat} 
\usepackage{subfig}

\usepackage{multirow} 
\usepackage{charter}

\begin{document}

\title{\bf Methodology for the use of proportional counters in pulsed fast neutron yield measurements}

\author{
Ariel Tarife\~no-Saldivia${}^{a,b}$\footnote{Corresponding author: atarifeno@cchen.cl, atarisal@gmail.com}, Roberto E. Mayer${}^{c,d}$, Cristian Pavez${}^{a,b}$, \\and Leopoldo Soto${}^{a,b}$\\
\textit{a) Comisi\'on Chilena de Energ\'ia Nuclear, Casilla 188-D, Santiago, Chile.}\\
\textit{b) Center for Research and Applications in Plasma Physics and Pulsed Power, $P^4$, Chile.}\\
\textit{c) Comisi\'on Nacional de Energ\'ia At\'omica, Univ. N. de Cuyo, Argentina.}\\
\textit{d) PLADEMA, Argentina.}
}
\date{}
\maketitle

\begin{abstract}
This paper introduces in full detail a methodology for the measurement of neutron yields and the necessary efficiency calibration, to be applied to the intensity measurement of neutron bursts where individual neutrons are not resolved in time, for any given moderated neutron proportional counter array. The method allows efficiency calibration employing the detection neutrons arising from an isotopic neutron source. Full statistical study of the procedure is descripted, taking into account contributions arising from counting statistics, piling-up statistics of real detector pulse-height spectra and background fluctuations. The useful information is extracted from the net waveform area of the signal arising from the electric charge accumulated inside the detector tube. Improvement of detection limit is gained, therefore this detection system can be used in detection of low emission neutron pulsed sources with pulses of duration from nanoseconds to up. The application of the methodology to detection systems to be applied for $D-D$ fusion neutrons from plasma focus devices is described. The present work is also of interest to the nuclear community working on fusion by magnetic confinement and lasers, and working on neutron production by accelerators or similar devices. 
\end{abstract}

\begin{flushleft}
\end{flushleft}

\section{Introduction}\label{intro}
Foil activation is usually used as a standard technique for neutron yield measurements in pulsed fusion sources. From the early plasma focus (PF) research \cite{mather1964,filippovetal1962},
activation of Silver and Indium foils have been used \cite{lanteretal1968, stephensetal1958, gentilinietal1980}, although the so called silver activated Geiger counter is the most known detector from both. In PF devices \cite{soto2005}, depending on the filling gas, neutrons from $D-D$ reactions or $D-T$ reactions are produced with typical energies of $2.45 MeV$ and $14.1 MeV$ respectively. For fast neutrons, activation detectors usually require neutron moderation to thermal energies. Notwithstanding, when neutron intensity is high enough, typically higher than $10^8 n$ per source pulse, activation by fast neutrons could be used as in the case of the Indium or Beryllium counters. \cite{gentilinietal1980, rowlandetal1984, mahmoodetal2006}. Detectors based on activation by thermal neutrons have usually detection limits higher than $10^5 n$ per source pulse \cite{lanteretal1968,gentilinietal1980,morenoetal2010}. 

In recent years, several groups around the world have started research programs on small scale low energy plasma focus devices, especially in the sub-kiloJoule energy range \cite{shyametal1978, sotoetal2001, dubrovskyetal2001,silvaetal2003, milaneseetal2003, routetal2008, vermaetal2008_2,krishnanet2009,el-aragi2010, velosoetal2011_correlations} and in the sub-hundredJoule energy range \cite{silvaetal2002,soto2005, sotoetal2008,sotoetal2009,tarifeno-saldiviaetal2011,sotoetal2010}. As a consequence, requirements on detector sensitivity were increased in order to characterize such devices. For the low emission regime ($Y<10^5 n$ per source pulse) it is possible to use moderated gas proportional counters based on ${}^3He$ or $BF_3$ filled tubes \cite{mayeretal1998}. In these detectors thermal neutron capture occurs with cross sections one order of magnitude higher than Indium or Silver activation cross section. Therefore, detection limits in neutron proportional counters should be at least one order of magnitude lower. The disadvantage of these detectors for pulsed sources  is the pulse piling-up produced during high instantaneous count rates due to the burst of fast neutrons, which make it impossible to count single neutron pulses by standard nuclear electronics. To deal with this problem two schemes of use have been proposed: \begin{itemize}
                                                                                                                                                                                                                                                                                                                                                                                                                                                                                                                                                                                                                                                                                                                                                                                                                                                                                                                                                                                                                                                                                                                                                                                                                                                                                                                                                                                                                                                                                                                                                                                                                                                                                                                                                                                                                                                                                                              \item[i)] Use of high counting rate electronics and controlling of the instantaneous count rate by detector-source separation. This technique has been reported to be successful in resolving yields higher than $10^6 n$ per source burst \cite{waheedetal2000}.

\item[ii)] Cross calibration on the accumulated charge (preamplifier output signal) generated in the counter tube by the neutron burst using as neutron reference a foil activation detector \cite{morenoetal2008, vermaetal2008_2}.
The comparison of the response of a reference detector with the proportional counter allows to obtain a calibration factor in the high emission regime. Thus, by means of extrapolation, it is possible to detect yields of the order of $10^3-10^2 n/burst$ in the low emission regime using the moderated proportional counters. Due to the fact that detection limits are transferred by the calibration process from the reference detector, the uncertainties in single measurements are of the order of the yield or greater. Thus, no improvement in measurement uncertainties is obtained for the low emission regime despite the high efficiency of neutron proportional counters tubes in comparison with foil activation detectors. However, for bursts of the order of $10^3  n/burst$, it is still possible to measure the pulsed neutron emission rate with an accuracy of the order of $40\%$ by taking averages on large samples of shots \cite{sotoetal2008}.                                                                                                                                                                                                                                                                                                                                                                                                                                                                                                                                                                                                                                                                                                                                                                                                                                                                                                                                                                                                                                                                                                                                                                                                                                                                                                                                                                                                                                                                                                                                                                                                                                                                                                                                                                                                                                                   \end{itemize}

The current state of the art in the use of proportional counters for pulsed fast neutrons lacks a calibration technique to allow reducing detection limits below $10^5 n$ per source burst in single shot measurements. To overcome this deficiency, in the following, a methodology is proposed for the analysis of the detector output signal to find the number of piled-up individual detected events per burst and its uncertainties, while allowing for detector efficiency to be established through the measurement of neutrons arising from an isotopic neutron source. Thus, it is possible to obtain the real neutron yield. The detailed study of the pulse piling-up  statistics and other sources that influence the measurement process, namely  counting statistics and background, are described in what follows.
 

\section{Working principles for fast neutron counting through a moderated proportional counter tube}\label{WorkingPrinciples}

 \begin{figure}
 \centering
 \includegraphics[width=0.48\textwidth,keepaspectratio]{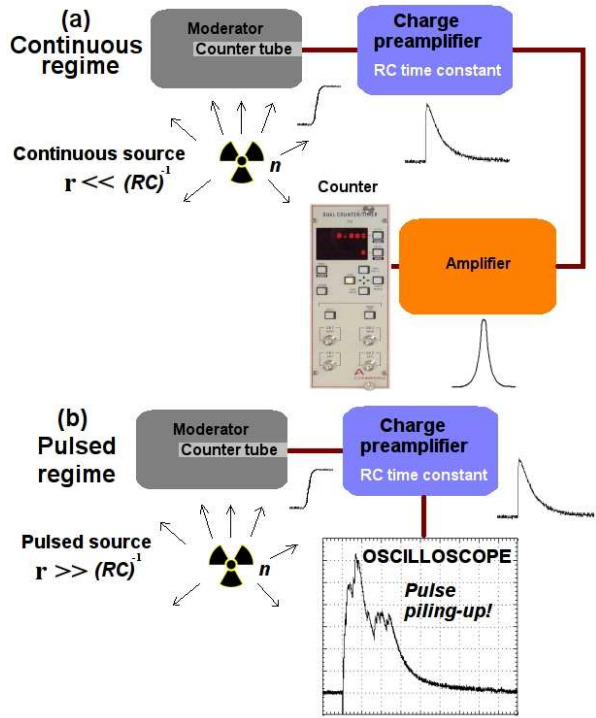}
\caption{Detection system setup and signal formation stages for (a) Continuous regime and (b) Pulsed regime.}
 \label{esq_uso_3he}
 \end{figure}

In the so called continuous regime, as depicted in fig. \ref{esq_uso_3he}, detection system consists of a moderated counter tube connected to a charge preamplifier and an amplifier, and finally connected to a counting system. As a consequence from the detection process in the counter tube, charge is produced and collected at the input terminal of the preamplifier. The output of the preamplifier is a signal whose pulse height is proportional to the integrated charge provided at the input. Decay of the preamplifier output is characterized by an exponential fall and $RC$ time constant. Once in the amplifier, typically a linear amplifier, the signal is shaped and amplified in order to make it suitable for counting systems. To allow proper counting, the count rate should in general satisfy $r<<(RC)^{-1}$. In fig. \ref{esq_uso_3he} it is also shown the output in the so called pulsed regime. In this case the preamplifier output is directly connected to a digital oscilloscope. When the detection system is irradiated by an extremely short pulse of fast neutrons ($ \tau \lesssim 10^2 ns$), the pulse is spread over tens of microseconds as a result of neutron moderation to thermal energies. When fluence is high enough to make the instantaneous count rate to be $r_{ins}>>(RC)^{-1}$,  single neutron signals will be piled-up thus smoothing the output signal and  making it impossible to count individual detected events by standard nuclear electronics. In contrast for lower fluence, when $r_{ins} \lesssim (RC)^{-1}$, signal shape is not smooth and eventually it is possible to distinguish single neutron pulses from the output signal. In both cases notwithstanding, the total number of detected events should be proportional to the output signal area. Even in the case of pulse piling-up, the number of detected events could be always obtained only from the net signal area, provided that space charge accumulation in the detector tube does not impair linearity.

When detection system is set up for pulsed regime (\emph{``charge integration mode''}), there are three sources of fluctuations that influence the measurement process:
\begin{enumerate}
 \item[i)] \textbf{Counting statistics:} Fluctuations arise from the detection process and are well described by the Poisson distribution.
 \item[ii)] \textbf{Pulse piling-up statistics:} Arise from the ``\emph{wall effect}'' taking place inside the counter tube, which results in a very asymmetrical distribution. Therefore, when pulse piling-up happens, the sum of two or more single random detected events generate overlapping distributions which give rise to differences between the true and the expected number of detected events.
\item[iii)]\textbf{Background:} The extracted signal from the detection system corresponds to the total charge generated in the tube, which is equivalent to the integral of the waveform acquired by the oscilloscope. Thus, electrical noise acts as background. Besides electrical noise, natural background is always present although in  some cases it could be negligible.
\end{enumerate}

\section{Detection system and experimental setup}\label{setup}

Detection systems available in our facilities were reported previously by Moreno \emph{et al} \cite{morenoetal2008}. They are based on a paraffin wax moderator ($45 \times 15 \times 15 cm^3$) and a ${}^3$He tube (model LND 2523) sheathed by a $3mm$ thick lead sheet to stop x-rays and prevent their subsequent detection. The moderator is surrounded by an external cadmium sheet to absorb thermal neutrons. The tube is connected to a preamplifier (CANBERRA 2006) and the high voltage feed is set to $4kV$. In order improve the signal-to-noise ratio, the preamplifier is adjusted with a conversion gain of $235 mV/M-ion-par$ which produces a $5\times$ voltage output factor scale \cite{preamp}. The $RC$ time constant is set to $50 \mu s$, this is the fall time for individual pulses when the input pulse risetime is less than hundred nanoseconds. When used for the pulsed regime, signal output from  the preamplifier is recorded by a digital oscilloscope (Tektronix model TDS 684) at $1 M\Omega$ input impedance, AC coupling, and $20MHz$ bandwidth. Horizontal scale is set in accordance with moderator decay time (governing neutron feeding into the ${}^3$He tube), in the case of our detectors time scale is set to $200 \mu s/div$. Direct measurements of the total signal area are obtained using the internal integration oscilloscope function on a $1500 \mu s$ time window, which starts close to the trigger point. The oscilloscope is triggered by the electromagnetic pulse generated from the electrical discharge. \\

Detector efficiency or calibration factor ($j_c$) for fast neutrons was obtained operating in the continuous regime by comparing neutron intensity from a standard isotopic source, used as neutron reference, and the registered net counting rate. The last is valid in the approximation that after many neutron scatters in the moderator the result is highly independent of the neutron input spectra; that is, the standard source spectra or that originated at the Plasma Focus. To satisfy this approximation the fast neutron reference should have a neutron energy spectra as similar to the fusion spectra as possible. Although $D-D$ reaction at low incident energies produces $2.45 MeV$ neutrons, observed fusion spectra around $2.45 MeV$ and extending from $1 MeV$ to $3.5 MeV$ have been reported in the literature \cite{schmidtetal1987,gentilinietal1980}. To fulfill this requirement a certified ${}^{252}$Cf isotopic source was used. This source is characterized by a mean energy between $2.1 - 2.3 MeV$ and its spectrum extends from $0.1-10 MeV$. Source neutron emision at the time of experiments was $6.5\times 10^4 s^{-1}$. Measurements of count rate were carried out at different positions from $14cm$ to $170cm$. For this purpose, the  preamplifier output was connected to a Tennelec TC-244 amplifier and then to a multichannel analyzer Camberra Multiport II controlled by the interface GENIE2K. The gross counting was corrected by discounting natural background and air attenuation according to that reported by Eisenhauer \emph{et al.} \cite{eisenhaueretal1985}. To account for solid angle, the neutron intensity was corrected by the source anisotropy factors \cite{eisenhaueretal1988}, which were previously characterized \cite{Atarifeno_PhdTesis}, and thus the number of neutrons arriving to the detector front face was calculated using the formula provided by Gotoh \emph{et al.} for a rectangular slit \cite{gotohetal1971,gotohetal1971erratum}. These same measurements of count rate provided the pulse height spectrum of each proportional counter. In the following sections results are referred to one of our detection systems which is called ${}^3$He-206 system.


\section{Piling-up statistics for proportional counters}\label{pilingupsts}
\begin{figure}
\centering
\includegraphics[width=0.45\textwidth,keepaspectratio]{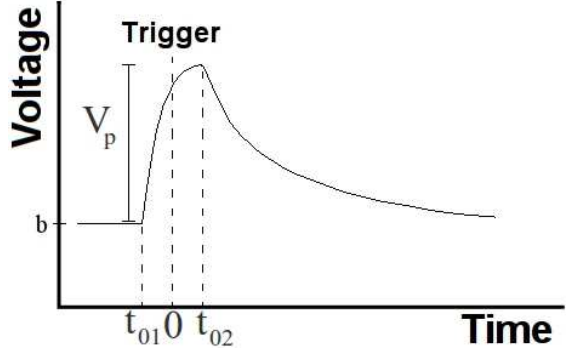}
\caption{Diagram of a typical output signal from the detection system associated to a single detected event.}
\label{fig:ModeloSenal}
\end{figure}
Piling-up statistics was studied by means of the acceptance-rejection Monte Carlo method, which was used to numerically construct  pulse piling-up distributions. The starting point was  to obtain the waveform area spectrum for single events. Due to pulse height at the preamplifier output being proportional to the accumulated charge provided at the input, then for single events waveform area is given by $A_p=V_p \cdot FA$, where $V_p$ is the pulse height and $FA$ a constant. Thus, the pulse area spectrum could be obtained from the pulse height spectrum through rescaling it by the constant $FA$. $FA$ depends only on the detection system. In order to characterize $FA$, let us first consider a typical output signal from the preamp associated to detection of a single event, as shown in fig. \ref{fig:ModeloSenal}. This signal can be modeled by the mathematical expression
\begin{equation}
\mathcal{S}(t) = \begin{cases}
 b \quad & \text{, if }t\leq -t_{01}.\\
 V_p(1-e^{-(t+t_{01})/\tau_1})+b \quad &\text{, if } -t_{01} \leq t \leq t_{02}.\\
V_p e^{-(t-t_{02})/\tau_2} +b \quad &\text{, if } t_{02}\leq t.\\
\end{cases}
\label{eq:ModeloSenal}
\end{equation}

Thus the signal area is given by
\begin{equation*}
	A_p = V_p \cdot (\tau_2 + \tau_1 (e^{-(t_{01}+t_{02})/\tau_1} -1 ) + t_{01}+t_{02}) 
\end{equation*}
where
\begin{equation}
	FA = (\tau_2 + \tau_1 (e^{-(t_{01}+t_{02})/\tau_1} -1 ) + t_{01}+t_{02}). 
\label{eq:factorArea2}
\end{equation}

\begin{figure}[tp]
\centering
\includegraphics[width=0.43\textwidth,keepaspectratio]{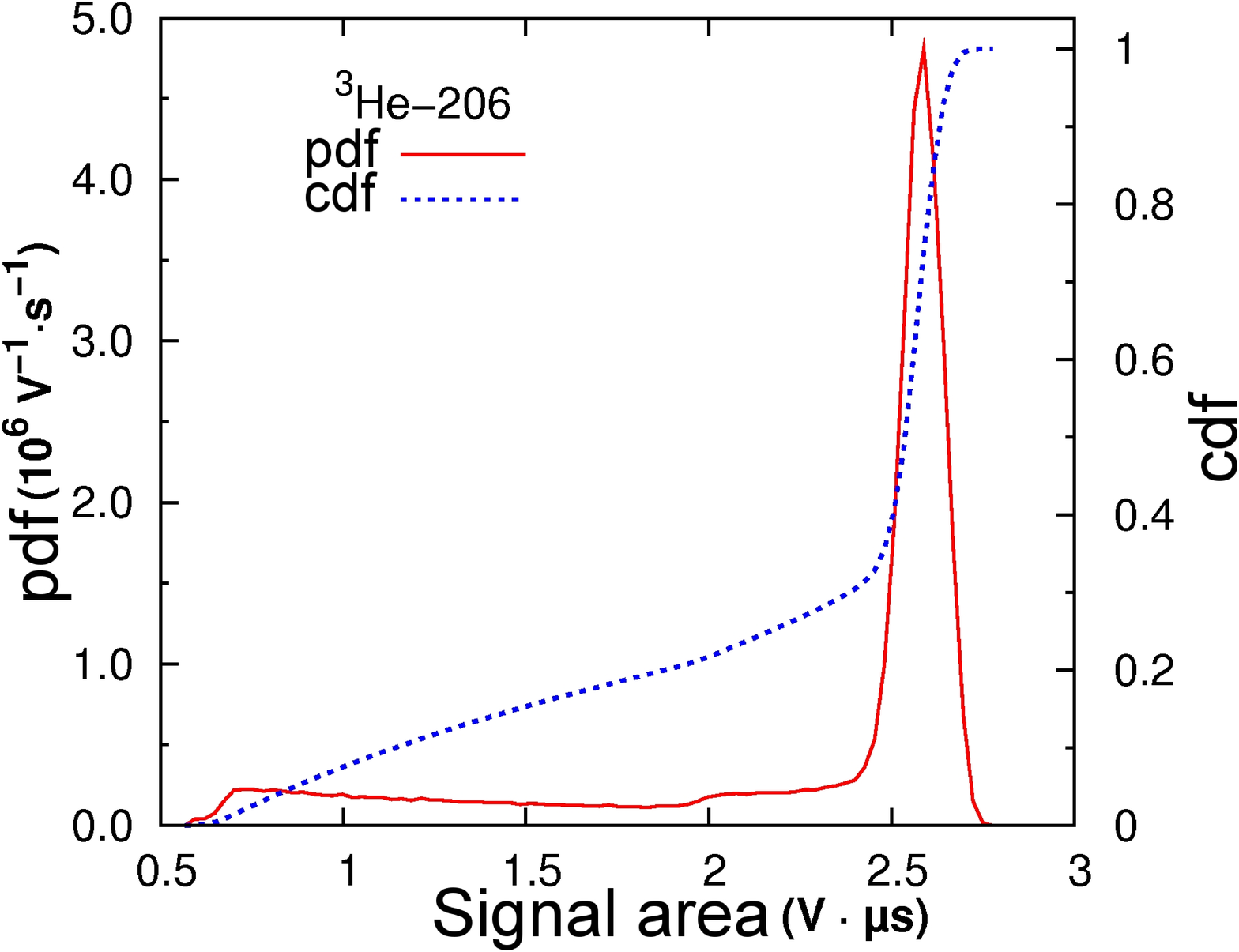}
\caption{Probability density function ($pdf$) and cumulative distribution function for the proportional counter tube in ${}^3$He-206 detection system.}
\label{fig:pdfcounter}
\end{figure}

To obtain proper values to calculate $FA$, hundreds of waveforms from single neutron events were analyzed using the model of eq. \ref{eq:ModeloSenal} and the non-linear least squares (Levenberg-Marquardt method) routine provided by the function \emph{fit} of the GNUPLOT package \cite{gnuplot}.  Results of this analysis are presented in table \ref{table:ResultsCaraterizacion_Senales}. Using these results it was possible to construct the probability density function (pdf) and the cumulative distribution function for the proportional counter tube as shown in figure \ref{fig:pdfcounter}.

\begin{table}[tp]
\begin{center}
\caption{Characteristic preamplifier parameters for single neutron events acording to eq. \ref{eq:ModeloSenal} in ${}^3$He-206 detection system.}
\begin{tabular}{ccccl}
\hline
\textbf{Parameter} & \multicolumn{4}{l}{\hspace{0.66cm} \textbf{Value}}\\ 
\hline
$\tau_1 (\mu s)$ & 1.24 &$\pm$& 0.01 &(0.5\%)  \\ 
$\tau_2 (\mu s)$ & 75.1 & $\pm$ & 0.3 & (0.4\%) \\ 
$t_{01}+t_{02} (\mu s)$ & 17.7 & $\pm$ & 0.1 & (0.6\%) \\ 
$FA(\mu s)$ & 91.6 & $\pm$ & 0.3 & (0.3\%) \\ 
\hline
\end{tabular}
\label{table:ResultsCaraterizacion_Senales}
\end{center}
\end{table}

Piling-up statistics was studied by simulating the random detector response in terms of the waveform area and the other statistics associated to the measurement process. Let $m$ be the total number of detected events and $x()$ a random number extracted from the counter tube $pdf$  (fig. \ref{fig:pdfcounter}) by means of the acceptance-rejection Monte Carlo method. When $m$ events are detected, the random ideal response of the detection system is given by
\begin{equation}
 X_T(m)=\sum_{j=1}^m x().
\label{ec:Resp_DetectorArea}
\end{equation} 
Thus the probability density function from the pulse piling-up statistics is obtained as a function of $m$ by sampling of equation \ref{ec:Resp_DetectorArea}. 

\begin{figure}[htp]
\centering
\includegraphics[width=0.37\textwidth,keepaspectratio]{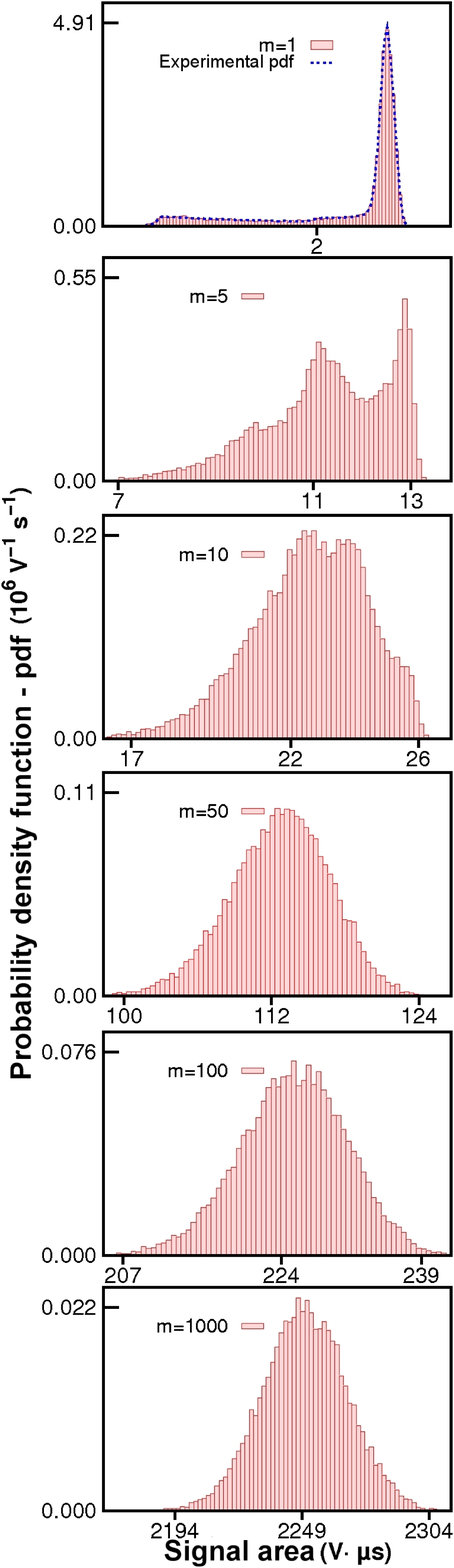}
\caption{Pulse piling-up distributions from sampling of eq. \ref{ec:Resp_DetectorArea} for ${}^3$He-206 detection system.}
\label{fig:pilingSts}
\end{figure}

Samples were generated numerically for $m:1,\ldots, 1000$. The sample size choice was $20000$. Results from the ${}^3$He-206 detection systems are shown in fig. \ref{fig:pilingSts}. As expected, for $m=1$ the numerical algorithm reproduces the experimental $pdf$ of the counter tube. From numerical results it is observed that when $m < 50$ pulse piling-up distribution is highly asymmetrical, for $m>100$ asymmetry is still present at tails of the distributions, finally when $m>500$ tail differences are not apparent from histograms. From the central limit theorem, the pulse piling-up statistic is expected to converge to the normal distribution. This property was evaluated at the $1\%$ significance level by means of Kolmogorov-Smirnov and Jarque-Bera normality test. Convergence of the  counter tube $pdf$ (fig. \ref{fig:pdfcounter}) to the normal distribution  was found in terms of cumulative probability (K-S test) when $m>90$, while in terms of asymmetry and peakedness (J-B test) when $m>800$.

\section{Background readings}\label{bkgn}
False positive readings are generated in the detection system by two mechanisms:
\begin{itemize}
\item[i)]\textbf{Natural background:} Corresponds to events from the environment which are characterized by a constant count rate. Natural background count rate in our experimental room is usually found in the range $0.3 - 1.0 s^{-1}$. These measurements were done under the continuous regime. 
For the pulsed regime the measurement time window is $1500 \mu s$, therefore contribution to the total number of detected events by natural background is negligible.

\item[ii)]\textbf{Electrical background noise (EBN):} In absence of detected events, the electrical noise contributes randomly to the total waveform area (detector reading), thus it behaves as a background reading. When the operative parameters in the detection system are fixed, readings from EBN respond to a normal distribution. The latter is concluded from analysis of EBN samples by the Kolmogorov-Smirnov test at $10\%$ significance level. 
\end{itemize}

These facts imply that the gross detector reading (${X_T}_{gross}$) should be corrected to discount background contributions. Let $B$ and $dB$ respectively be the sample mean and sample standard deviation associated to the EBN. Therefore, the net detector reading is obtained from 
\begin{equation}
	{X_T}_{net}={X_T}_{gross}-B.
\label{eq:areaNeta}
\end{equation}

For a specific time window, the EBN sample statistical parameters depend on oscilloscope horizontal and vertical scale, they are affected by changes in laboratory room temperature and should be monitored in-situ together with neutron measurements for shots with no neutron detection.  

\section{Counting model from the waveform area}\label{CountingModel}

To arrive at a counting model as a function of the net waveform area, let us first consider a sequence of random values associated to $m'$ detected events: $\lbrace {X_T}_1, {X_T}_2, \ldots, {X_T}_n\rbrace$. Therefore the following expressions are satisfied 
 \begin{eqnarray*}
m' &=& a_1 {X_T}_1\\
m' &=& a_2 {X_T}_2\\
&\vdots& \\
m' &=& a_n {X_T}_n.
\end{eqnarray*}
Selecting $a_i=a$ as a constant, let the following $m'$ be defined as follows,
\begin{eqnarray*}
m'_1 &=& a \cdot  {X_T}_1\\
m'_2 &=& a \cdot {X_T}_2\\
&\vdots& \\
m'_N &=& a \cdot {X_T}_n.
\end{eqnarray*}
By properly choosing $a$, convergence of the distributions ensures that
\begin{equation}
	E(m'_i)=E(a \cdot {X_T}_i)=a \cdot E({X_T}_i)=m'.
\end{equation}
Being $E(x)$ the mathematical expectation of the variable $x$. This constant $a$ was evaluated using samples from the pulse piling-up statistic. Detected number of events was plotted against the sample mean of the waveform area,
\begin{equation}
	m(E({X_T}_i))=a \cdot E({X_T}_i).
\label{ec:ajuste_a}
\end{equation}
Thus, a linear least squares through the origin yields the value of $a$. Results of this analysis are shown in table \ref{tab:LinearModels}.

\begin{table}[tbp]
\caption{Linear fit results using model from equation \ref{ec:ajuste_a}.}
\begin{center}
\begin{tabular}{cc}
\hline
\textbf{Parameter} &\textbf{ Value}  \\ 
\hline
{\bf Range}  & $m: 1-300$ \\ 
{\bf Sample size} & $20000$  \\ 
 $\boldsymbol \delta m$ & 1 \\ 
$\boldsymbol a (V^{-1} \cdot \mu s^{-1})$ & $0.44459 \pm 0.00001$ \\ 
{\bf fit} $\boldsymbol r^2$ & $0.99999907$ \\ 
\hline
\end{tabular}
\end{center}
\label{tab:LinearModels}
\end{table}   

Finally the counting model is defined by means of the rounding function, which round the argument to the nearest integer, as follows
\begin{equation}
	m_{mc}=G(X_T)=[a\cdot X_T],
\label{eq:defCountingModel}
\end{equation}
where $m_{mc}$ corresponds to a predicted value for the number of detected events from the signal waveform area  or accumulated charge. Once the counting model has been obtained, it becomes necessary to study the predictive power of the method and the uncertainties. As already stated in section \ref{WorkingPrinciples}, in practical situations fluctuations are due to the counting statistics, which responds to the Poisson distribution, also due to the pulse piling-up statistics whose distribution has been studied in section \ref{pilingupsts} and  to the electrical background noise which is governed by the normal distribution (see section \ref{bkgn}).  

Predictive power was studied by simulating random measurements as predictions from the counting model. For that purpose pseudo-random numbers from the Poisson and normal distributions where required. Due to its proven mathematical properties although being slower than others, the \verb|gsl_rng_ranlux389| pseudo-random number generator provided by the \emph{GNU Scientific Library} (GSL) was used \cite{GSL}. The algorithm for simulating random measurements and model predictions is described as follows:
\begin{enumerate}
\item Let $m$ be the true number of detected events.
\item Let $m_{ec}$ be a pseudo-random integer extracted from the Poisson distribution with expected value $m$.
\item Let ${X_T}_{ec}$  be a  pseudo-random number extracted from the pulse piling-up statistic associated to $m_{ec}$ detected events.
\item Let  $B_{ran}$  be a  pseudo-random number extracted from a Gaussian distribution ($\mu=0$) with $\sigma=dB$. Thus a new variable is defined  by ${X_{T}}_{net}={X_T}_{ec}+B_{ran}$. At this point,  ${X_{T}}_{net}$ corresponds to the simulated net detector reading.
\item Finally a prediction from the counting model is obtained by $m_{mc}=G({X_T}_{ec})$.
\end{enumerate}

Sampling of the above described algorithm allows further calculation of the statistical parameters $E(m_{ec})$,  $E(m_{mc})$,  $E(B_{ran})$, $\sigma(m_{ec})$, $\sigma(m_{mc})$  and $\sigma(B_{ran})$, and comparison of them with the true number of detected events. 

Samples were created for the interval $1\leq m <300$ with a step size of $\delta m=2$. Sample size choice was $5000$. For each value of $m$, $dB$ was varied in the interval $0 \leq dB \leq 10 ~ V\cdot \mu s$ with steps of $\delta dB=0.25 V\cdot \mu s$ when $dB\leq 2.0 V\cdot \mu s$, and $\delta dB=0.5 V\cdot \mu s$ when $2.0< dB\leq 10 V\cdot \mu s$.

Analysis of numerical results to determine predictive power and model uncertainties was done by means of linear models and hypothesis testing. For predictive power, if predictions from the counting model ($E(m_{mc})$) converge to the true number of detected events ($m$), then these two variables should be related in a $X-Y$ plot by the identity function. To evaluate shift of the counting model caused by its definition or by numerical reasons, a general linear model is used: $y(x)=\beta_1 x + \beta_0$. Thus, the hypothesis to be tested by the F-test is given as follows:

\begin{table}[h]
\begin{center}
 \begin{tabular}{cccl}
$H_0:  \beta_1 = 1 $ &y & $\beta_0 = 0 $& Null hypothesis\\
& & & \\
& \emph{versus}& &\\
\multirow{2}{*}{$H_a:  \beta_1 \neq 1$} & \multirow{2}{*}{y}& \multirow{2}{*}{$\beta_0 \neq 0$}  & Alternative\\
& & & hypothesis 
\end{tabular}
 \end{center}
\end{table}

If null hypothesis is accepted for a given significance level $\alpha$, then a high predictive power is concluded for the counting model. Otherwise, the counting model is assumed to be poorly predictive and should be rejected. 

\begin{figure}[tp]
\centering
\includegraphics[width=0.48\textwidth,keepaspectratio]{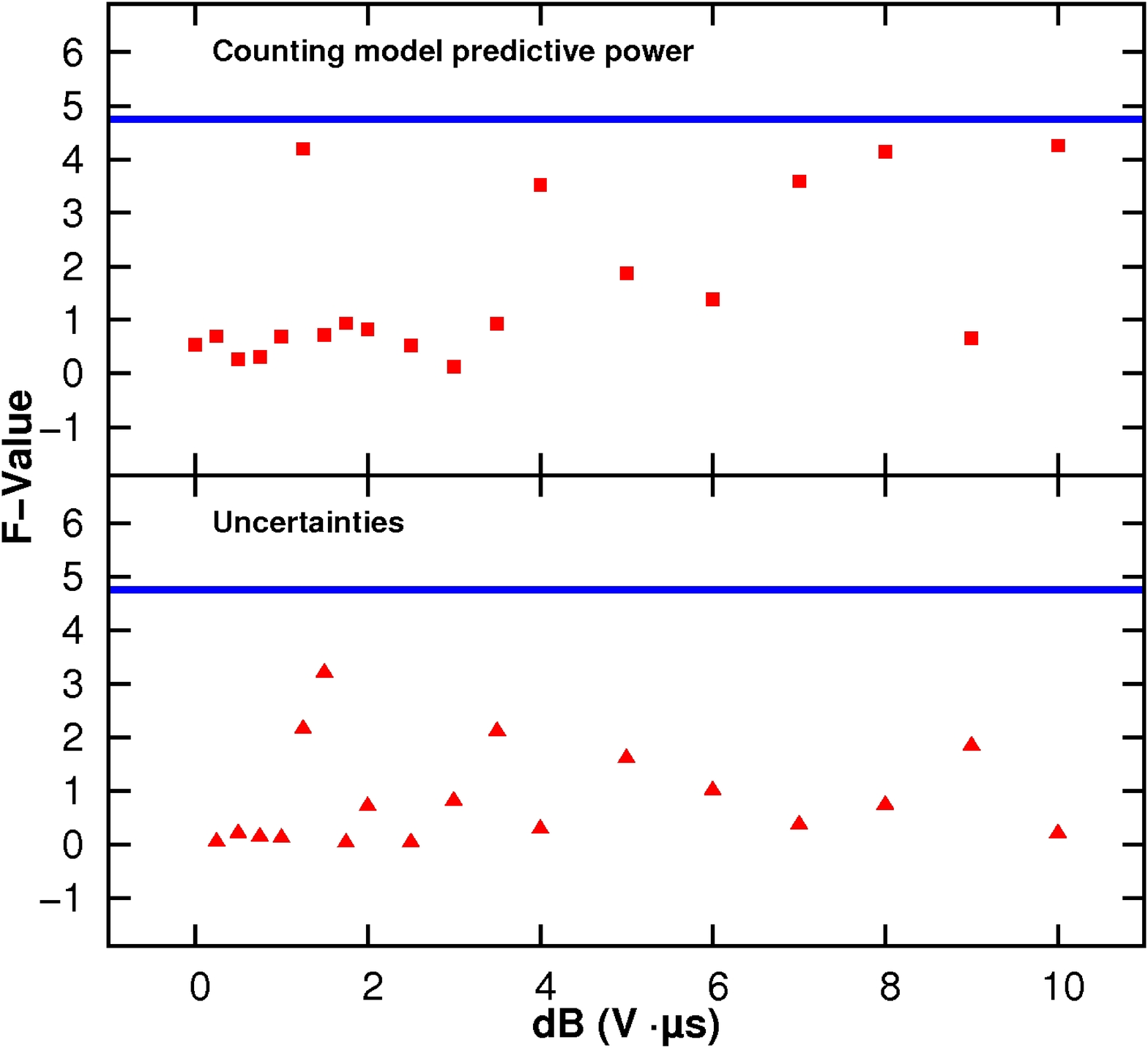}
\caption{Test of hypothesis results for predictive power of counting model and uncertainties. Critical value for the F-test at the $1\%$ significance level is shown as a solid line.}
\label{fig:predictivepower}
\end{figure}

For the F-test, the significance level was chosen to be $1\%$ ($\alpha=0.01$). Therefore, our sample with $148$ degrees of freedom has a F-test critical value of $4.751$. At the top of figure  \ref{fig:predictivepower} results are shown  of application of the F-test on samples as a function of $dB$ for the study of counting model predictive power. For each sample, it was verified that the expected value of the counting statistics also converge to the true number of detected events. From figure  \ref{fig:predictivepower}, it is observed that all calculations are less than the critical F-value, thus the counting model defined by equation \ref{eq:defCountingModel} is concluded to be highly predictive, this is to say
\begin{equation}
 E(m_{mc})=m.
\label{eq:conclModelPredPower}
\end{equation}
The last being valid at least in the interval under study: $m<300$ and $dB \leq 10 V\cdot \mu s$. 

Once counting model is accepted as predictive, it is necessary to study uncertainties of the model. With this purpose, two cases were considered:
\begin{itemize}
 \item \textbf{Case 1:} Uncertainties of the counting model due only to counting statistic and pulse piling-up statistic. This case is equivalent to consider sample generation when 
\begin{equation}
 \lim_{dB \rightarrow 0} B_{ran} \longrightarrow 0.
\end{equation}

\item \textbf{Case 2:} Uncertainties of the counting model due to counting statistic, pulse piling-up statistic and electrical background noise.

\end{itemize}

For case 1 samples in the interval $1\leq m \leq 300$, with $\delta m=1$ and sample size of $20000$, were analyzed. As a result, it was found that variance of predictions from the counting model are proportional to the expected value for predictions from the counting model,
\begin{equation}
\sigma^2(m_{mc})=  f_0\cdot E(m_{mc})=f_0\cdot m.
\label{eq:modeloConteo_DesvSTD}
\end{equation}  

\begin{figure}[htp]
\centering
\includegraphics[width=0.48\textwidth,keepaspectratio]{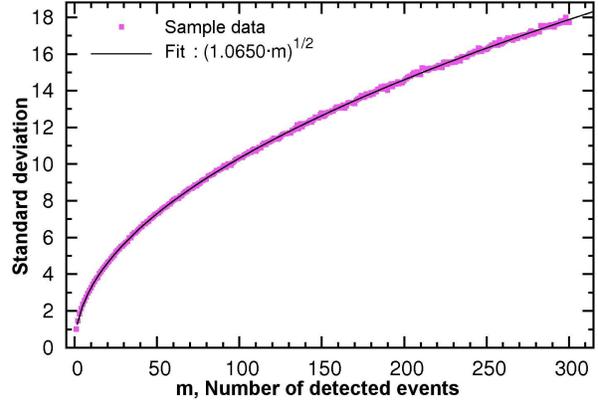}
\caption{Standard deviation modeling from sample data for ${}^3$He-206 detection system.}
\label{fig:stdDeviations}
\end{figure}

Thus, distributions generated by the counting model, neglecting background effects, could be said to be \emph{Poisson-like} because effects of the counting statistic dominate over pulse piling-up statistic. Last conclusions are supported by numerical results as shown in figure \ref{fig:stdDeviations} and table \ref{tab:Ajustes_Desvest_ModConteo}. On one hand, samples from the counting statistics were analyzed by linear least squares (LLS) using the model $\sigma(m_{ec})^2=\gamma \cdot m$. Use of the T-test of hypothesis allowed to conclude for the $1\%$ significance level that $\gamma=1$, as expected. On the other hand, analysis of sample data by LLS and model results from equation \ref{eq:modeloConteo_DesvSTD} allow to conclude that $f_0 =1.065$. Therefore, it is possible to say that pulse piling-up statistical contribution is $6.5\%$ the contribution from the counting statistic to counting model variance when neglecting background effects.

\begin{table}[htbp]
\caption{Counting model uncertainties study for case 1.}
\begin{center}
\begin{tabular}{ccc}
\hline
 & \textbf{Parameter} & \textbf{Value}  \\ \hline

Counting  & $f_0$ & $\boldsymbol{1.065\pm0.0006}$ \\ 
model & $r^2$ & $0.9995687$   \\ 
\hline
Counting & $\gamma$ & $0.9998\pm0.0006$ \\ 
statistic &  $r^2$ & $0.99958056$ \\
\hline
\end{tabular}
\end{center}
\label{tab:Ajustes_Desvest_ModConteo}
\end{table}

For case 2 is postulated that counting model variance is given by, 
\begin{equation}
\begin{split}
 \sigma^2(m_{mc})&=f_0\cdot E(m_{ec}) + \sigma^2(a\cdot B_{ran})\\
&=f_0\cdot m + a^2 \cdot dB_{sample}^2.
\end{split}
\label{eq:fluctuaciones_ModeloConteo}
\end{equation}

To demonstrate the last hypothesis, predictive power of equation \ref{eq:fluctuaciones_ModeloConteo} was studied. If the right side of equation \ref{eq:fluctuaciones_ModeloConteo} converges to the left one, then the plot $\sigma^2(m_{mc}) \quad v/s  \quad f_0\cdot m + a^2 \cdot dB_{sample}^2$ should be given by the identity function. By considering samples with the same characteristics than those set up for the study of counting model  predictive power, it was obtained that equation \ref{eq:fluctuaciones_ModeloConteo} is highly predictive for the sample counting model standard deviation, as it could be verified at bottom of figure  \ref{fig:predictivepower}. It is important to note that to achieve predictability, the sample standard deviation of the background electrical noise should be used.   

Summarizing results so far, it has been demonstrated that definition of the counting model by equation \ref{eq:defCountingModel} ensures  $E(m_{mc})=m$ and $ \sigma^2(m_{mc})= f_0\cdot m + a^2 \cdot dB_{sample}^2$.

\section{Neutron yield measurements}\label{ssec:NYield}

\begin{figure}[htp]
\centering
\includegraphics[width=0.48\textwidth,keepaspectratio]{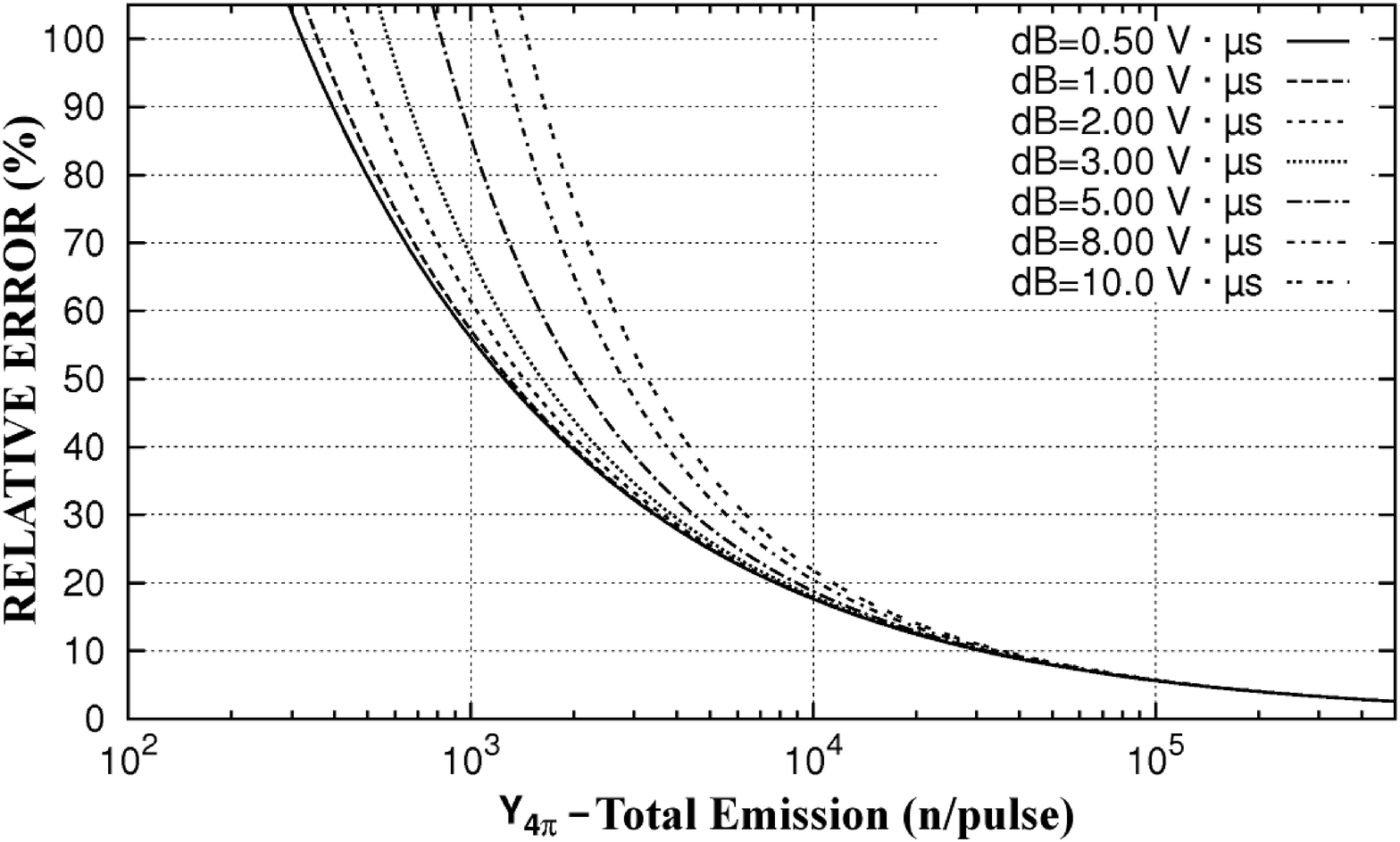}
\caption{Mean relative error for ${}^3$He-206 detection system.}
\label{fig:relativeError}
\end{figure}

Let us consider a single shot from the pulsed source, e.g a plasma focus device. Now we deal with the problem of estimating the expected number of detected events and its uncertainties. Let ${X_T}_{net}$ the net detector response (see equation \ref{eq:areaNeta}). Then, for this single measurement the best estimation of the expected number of detected events comes from the measurement itself \cite{knoll1989},
\begin{equation}
 m = E(m_{mc}) \cong G({X_{T}}_{net}) = [a\cdot {X_{T}}_{net}].
\label{eq:EstSingleMeas}
\end{equation}
Thus, the uncertainty associated to the expected number of detected events in a single measurement is calculated from 
 \begin{equation}
 \begin{split}
  \sigma(m)&=\sqrt{f_0\cdot m + a^2 \cdot dB_{sample}^2}\\
& \cong  \sqrt{f_0\cdot G({X_{T}}_{net}) + a^2 \cdot dB_{sample}^2}\\
&=\sqrt{f_0\cdot [a\cdot {X_{T}}_{net}] + a^2 \cdot dB_{sample}^2}
 \end{split}
\label{eq:EstUncerSingleMeas}
\end{equation}

Once obtained an estimation of the number of detected events has been obtained, the neutron yield is easily deduced by using the calibration factor of the detection system ($j_c$). Consequently, the number of neutrons per solid angle is given by
\begin{equation}
 \begin{split}
  Y= &j_c\cdot [a\cdot {X_{T}}_{net}] \cdot \\
&\left( 1 \pm  \sqrt{\left( \frac{dj_c}{j_c} \right)^2 +   \frac{f_0}{[a\cdot {X_{T}}_{net}]} + \frac{a^2 \cdot dB_{muestral}^2}{[a\cdot {X_{T}}_{net}]^2}}   \right).
 \end{split}
\label{eq:estimacionYield}
\end{equation}

Let $\Omega$ be the solid angle subtended by the detector at its measurement position, then $f_{4\pi}=4\pi/\Omega$. Thus under the assumption of isotropic emission, the total neutron emission per pulse is given by 
\begin{equation}
 Y_{4\pi}=f_{4\pi} \cdot Y.
\end{equation}

\begin{table}[tbp]
\begin{center}
\caption{Characteristic values for $j_c$ and $f_{4\pi}$ in the interval $14 cm\leq r \leq 40cm$.}
\begin{tabular}{cc}
\hline
\textbf{Parameter}& \textbf{Mean}\\
\hline
$\bar j_c$  & $14.2 \pm 0.1$\\
$\bar f_{4\pi}={4\pi}/\bar \Omega$ & $20.45$ \\ \hline
\end{tabular}
\label{tab:ValCarac_rangoPos}
\end{center}
\end{table}

In order to study detection limits, let us consider table \ref{tab:ValCarac_rangoPos} where characteristic values for $j_c$ and $f_{4\pi}$ of the typical positioning interval in an extremely low energy plasma focus device (PF-50J, \cite{sotoetal2008}) are shown.  In addition, from $ [a\cdot {X_{T}}_{net}]= Y/\bar j_c $ and  $Y= Y_{4\pi}/ \bar f_{4\pi}$ the standard relative error is then calculated  by 
\begin{equation}
 \begin{split}
  \% &dY (Y)= 100\cdot\\
& \sqrt{\left( \frac{d\bar j_c}{\bar j_c} \right)^2 +   \frac{\bar f_0 \cdot {\bar j_c}\cdot \bar f_{4\pi}}{Y_{4\pi}}+\left( \frac{\bar a\cdot dB_{sample} \cdot {\bar j_c} \cdot \bar f_{4\pi}}{Y_{4\pi}}  \right)^2}.
 \end{split}
\label{eq:final_porcError}
\end{equation}

Plots from equation \ref{eq:final_porcError} are shown as a function of $dB_{sample}$ in figure \ref{fig:relativeError}. It is observed from this figure that when total neutron emission is higher than $10^5 n/pulse$, uncertainties are less than $10\%$. For $10^4 n/pulse$ uncertainties are of the order $20\%$.  Finally, it has to be pointed out that reasonable uncertainties ($<40\%$) are still obtained in the range $2\cdot 10^3<Y_{4\pi}<10^4 n/pulse$. The last is a direct consequence of the methodology proposed in this work.

\section{Concluding remarks}\label{conclusion}
The statistics of the signal generated from pulse piling-up (pulse piling-up statistics) in a neutron proportional counter when used in the herein called pulsed regime has been studied. It has been found that pulse piling-up statistics converge to a normal distribution typically when the number of piled-up pulses is higher than a hundred. 
From the pulsed piling-up statistics, the counting statistical behaviour and background contributions, a measurement methodology for bursts of neutrons has been developed. The method is based on a counting model for the number of detected events  based upon the net accumulated charge at the output of the detection system (waveform area). It has been demonstrated that the model, as defined in equations \ref{eq:defCountingModel} and \ref{eq:modeloConteo_DesvSTD}, ensures   $E(m_{mc})=m$ and $ \sigma^2(m_{mc})= f_0\cdot m + a^2 \cdot dB_{sample}^2$, when $m$ is the true number of detected events. Then by using equation \ref{eq:estimacionYield} it is possible to calculate the neutron emission, bearing in mind that the model holds while space charge accumulation in the detector tube does not run so high as to impair linearity. By this methodology, it has been possible to achieve detection limits almost two orders of magnitude lower than those from state of the art techniques. The methodology is quite general and could be reproduced by other groups in order to calibrate neutron detectors based on moderated proportional counters for fast neutrons.

\section*{Acknowledgments}
This work was supported by the Chile Bicentenial Program in Science and Technology grant ACT 26, Center for Research and Applications in Plasma Physics and Pulsed Power Technology (P4 Project).






\bibliographystyle{elsarticle-num}







\end{document}